\documentclass[iop]{emulateapj}
       \usepackage{graphicx}
       \usepackage{natbib}
\usepackage{graphicx,rotating}
\usepackage{amssymb}
\usepackage{epstopdf}
\usepackage{amsmath}
\usepackage{natbib}
\usepackage{xcolor}
\usepackage{xspace}
\newcommand{\beq}{\begin{equation}}
\newcommand{\eeq}{\end{equation}}
\def\etal{{\sl et~al.~}}
\newcommand{\Hers}{{14\,Her~}}
\newcommand{\Her}{14\,Her}

\newcommand{\HST}{{\it HST}}
\newcommand{\HSTs}{{\it HST~}}
\newcommand{\FGS}{{\it FGS}}
\newcommand{\FGSs}{{\it FGS~}}
\newcommand{\HIP}{{\it Hipparcos}}
\newcommand{\G}{{\it Gaia}}
\newcommand{\Gs}{{\it Gaia~}}

\newcommand{\ms}{m s$^{-1}$}

\newcommand{\msini}{$\cal{M}$\,sin\,{i}~}

\newcommand{\my}{mas yr$^{-1}$}
\newcommand{\m}{$\cal{M}$}
\newcommand{\mjup}{$\cal{M}_{\rm Jup}~$}
\newcommand{\mjupe}{$\cal{M}_{\rm Jup}$}
\newcommand{\msun}{$\cal{M}_{\odot}~$}
\newcommand{\msune}{$\cal{M}_{\odot}$}
\newcommand{\cosr}{\rm cos}
\newcommand{\sinr}{\rm sin}

\received{}
\revised{} 
\accepted{}

\shorttitle{The \Hers System}
\shortauthors{Benedict \etal}

\begin{document}
\bibliographystyle{/Users/gfbenedict/Active/my2}

\title{The \Hers Planetary System: Companion Masses and Architecture from Radial Velocities and Astrometry}
\author{G. F. Benedict}
\affiliation{McDonald Observatory, University of Texas, Austin, TX 78712}
 \author{B. E.  McArthur}
\affiliation{McDonald Observatory, University of Texas, Austin, TX 78712}
\author{E. P. Nelan}
\affiliation{Space Telescope Science Institute, Baltimore, MD 21218}
\author{J. L. Bean}
\affiliation{Department of Astronomy \& Astrophysics, University of Chicago, Chicago, IL 60637 }

\begin{abstract}
We combine {\it Hubble Space Telescope} (\HST) Fine Guidance Sensor, \HIP, and \Gs DR3 astrometric observations of the K0 V star 
\Hers  with the results of an analysis of extensive ground-based radial 
velocity  data to determine  perturbation orbits and masses  for two previously known companions, \Her\,b and c.     
Radial velocities obtained with the Hobby-Eberly Telescope and from the literature  now span over twenty five years. With these data we obtain  improved RV orbital elements for both the inner companion, \Her\,b and the long-period outer companion, \Her\,c.  We also find  evidence of an additional RV signal with P$\sim3789^{\rm d}$. 

We then model astrometry from \HIP, \HST, and \Gs  with RV results
 to obtain system parallax and proper 
motion, perturbation periods,  inclinations, and  sizes due to \Her\,b and c. We find  
$P_{\rm b}$ = 1767.6 $\pm$ 0.2 d,  perturbation semi-major axis $\alpha_{\rm b} =1.3 \pm 0.1$ mas, and  
inclination $i_{\rm b} = 36\arcdeg \pm 3\arcdeg$, $P_{\rm c}$ = 52160 $\pm$ 1028 d,  perturbation semi-major axis $\alpha_{\rm c} =10.3 \pm 0.7$ mas, and  
inclination $i_{\rm c} = 82\arcdeg \pm 14\arcdeg$.   In agreement with a past investigation, the \Her\,b, c orbits exhibit significant mutual inclination. Assuming a primary mass 
$M_* = 0.98\pm0.04 M_{\sun}$, we obtain  companion masses \m$_{\rm b} = 8.5 ^{ +1.0}_{-0.8} $\mjup and  \m$_{\rm c} = 7.1 ^{ +1.0}_{-0.6}$\mjupe.   

\end{abstract}


\section{Introduction}
Exoplanetary systems provide an opportunity  to probe the dynamical origins of planets \citep[e.g.][]{For06} and system evolution \citep{Wri09}.  They provide laboratories within which to tease out the essential processes and end states from the accidental.  The nearby, metal-rich  KO V star, \Hers (HD 145675), has long been known to host a companion \citep{Butl03}, and likely hosts a second \citep{Goz06, Wit07, Wri07, Hir21, Ros21}. Recent astrometric and radial velocity (RV) explorations of the \Hers multi-planet system include \cite{Bar21} and  \cite{Fen22}. 

Before it was included in lists of multi-planet systems, we included \Hers in a {\it Hubble Space Telescope} (\HST) proposal \citep{Ben05p} to carry out astrometry using the Fine Guidance Sensors (\FGS). Those observations supported attempts to establish true component mass of several promising candidate systems, all relatively nearby, and with companion \msini values and periods suggesting measurable astrometric amplitudes. See \cite{Ben17}, section 4.6, for a review.  As discussed below, these \Hers \FGSs data were problematical, and until now, not fully analyzable.

 We now return to these older \FGSs data, motivated by newer  predictive resources. These include the \Gs DR3 $RUWE$ parameter which  predicts unmodeled photocenter motion \citep{Sta21}, and the \cite{Bra21} $\chi^2$ value. The latter parameter measures an amount of measured acceleration obtained by comparing an earlier epoch proper motion from \HIP~with a DR3 proper motion. A larger $\chi^2$ value indicates more significant  change (acceleration) in proper motion, thus a higher probability of a perturbing companion. For \Hers $\chi^2 = 1009$ and $RUWE=1.819$, both indicating significant difference from straight line motion.

\Hers is a system for which accurate component masses would prove useful. For \Hers we use models previously employed for the exoplanet candidate systems $\epsilon$ Eri \citep{Ben06}, HD 33636 \citep{Bea07}, $\upsilon$ And \citep{McA10}, HD 136118 \citep{Mar10}, HD 38529 \citep{Ben10}, HD 128311 \citep{McA14}, HD 202206 \citep{Ben17b}, and $\mu$ Ara \citep{Ben22c}. 

A mass for \Her\,b  was our original goal. Returning to this target, now with a suspected second companion,  raises the possibility of establishing two companion masses and system architecture.

Section~\ref{Ap1} identifies the sources of RV data and our RV modeling results for components \Her\,b and c. Section~\ref{AST} describes the astrometric data and modeling techniques used in this study, now enabled by reference star astrometry from the \Gs DR3 catalog, and ICRS positions for \Hers from \HIP, \Gs DR1, and DR3. Including  orbits for \Her\,b and c, informed by orbital priors from the RV analysis,  yields orbital elements and  masses (Section~\ref{ORBIT}).  We compare our results with previous investigations, and place them in the context of  past FGS astrometric results (Section~\ref{Disc}). 
Lastly, in Section~\ref{Summ} we summarize our findings. 
For this investigation we adopt the \Hers stellar properties presented in table 1 of \cite{Bar21}, most critically the host star mass, \m$_\star=0.98\pm0.04$\msune.  For our assessment of the reality of candidate companion \Her\,d, we also adopt from their table 1 an activity parameter, $logR'_{HK}=-4.94\pm0.04$, a rotation period, $P_{rot}=29.5^{\rm d}$, and an age estimate of $4.6^{+3.8}_{-1.3}$Gyr.

We abbreviate millisecond of arc as mas throughout and state times as mJD=JD-2400000.5. 

\section{Radial Velocities}  \label{Ap1}

Sources (Table~\ref{tbl-RV}) of RV data include previously published ELODIE \citep{Nae04}, the Automated Planet Finder (APF) at Lick Observatory \citep{Vog14}, and Keck HIRES \citep{Butl03,Wit07,Hir21,Ros21}. The Hobby-Eberly Telescope High Resolution Spectrograph \citep{Tul98} provided previously unpublished measurements, produced using
the pipeline described in \cite{Bea07}.

All four sources of relative RV have differing zero-points. Our modeling includes zero-point offsets as solved-for parameters, listed  in Table~\ref{tbl-RV}, where we have arbitrarily assumed a 0.0 zero-point for the Keck data. Because our GaussFit \citep{Jef88} modeling results critically depend on the input data errors, we first modeled the RV to assess the validity of the original input RV errors.  In order to achieve a $\chi^2/DOF$ of unity for our solution required increasing the original errors on the RV  by a  factor of 2.9 for all sources. This suggests that either the errors were underestimated, or that  that the fit is not as good as it could be (i.e. evidence that there may be more to learn about the system), or a level of stellar-induced RV noise. 

\begin{figure}
\includegraphics[width=6.5in]{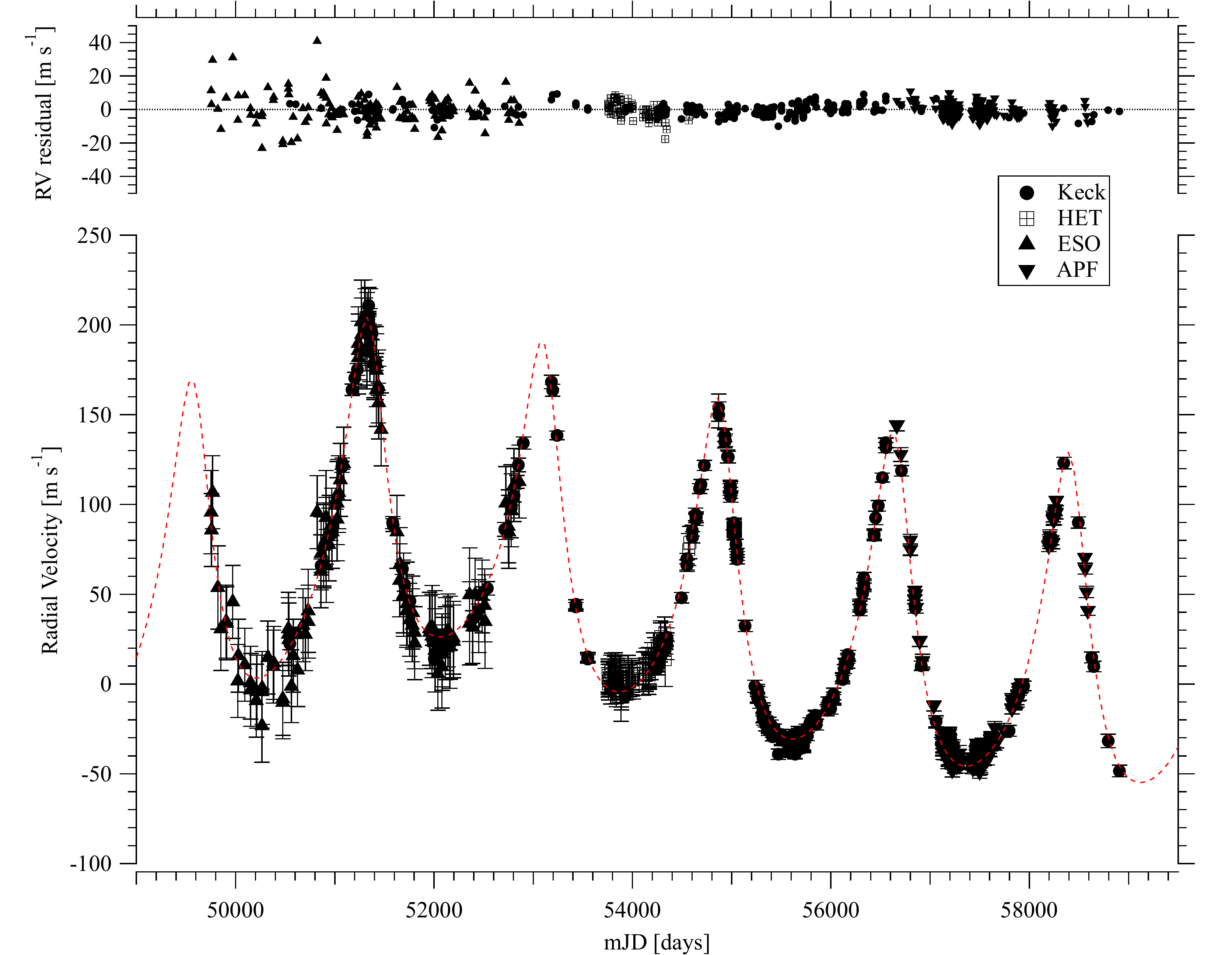}
\caption{RV values and $1-\sigma$ errors from the sources listed in Table~\ref{tbl-RV} plotted on the final RV two component orbit (Table~\ref{tbl-RVorb}). All RV input errors have been increased by a factor of 2.9 to achieve a near unity $\chi^2/DOF$. Residuals are plotted in the top panel. We note the RMS RV residual values for each source in Table~\ref{tbl-RV}.} \label{fig-RVf}
\end{figure}

Figure~\ref{fig-RVf} presents RV plotted as a function of time and the final combined orbital solution. Table~\ref{tbl-ALLRV} lists the final zero-point corrected RV, along with our scaled errors. Table~\ref{tbl-RV} also includes the RMS residual for each source. Table~\ref{tbl-RVorb} contains orbital elements and $1-\sigma$ errors for \Her\,b,c. 

\begin{center}
\begin{figure}
\includegraphics[width=4in]{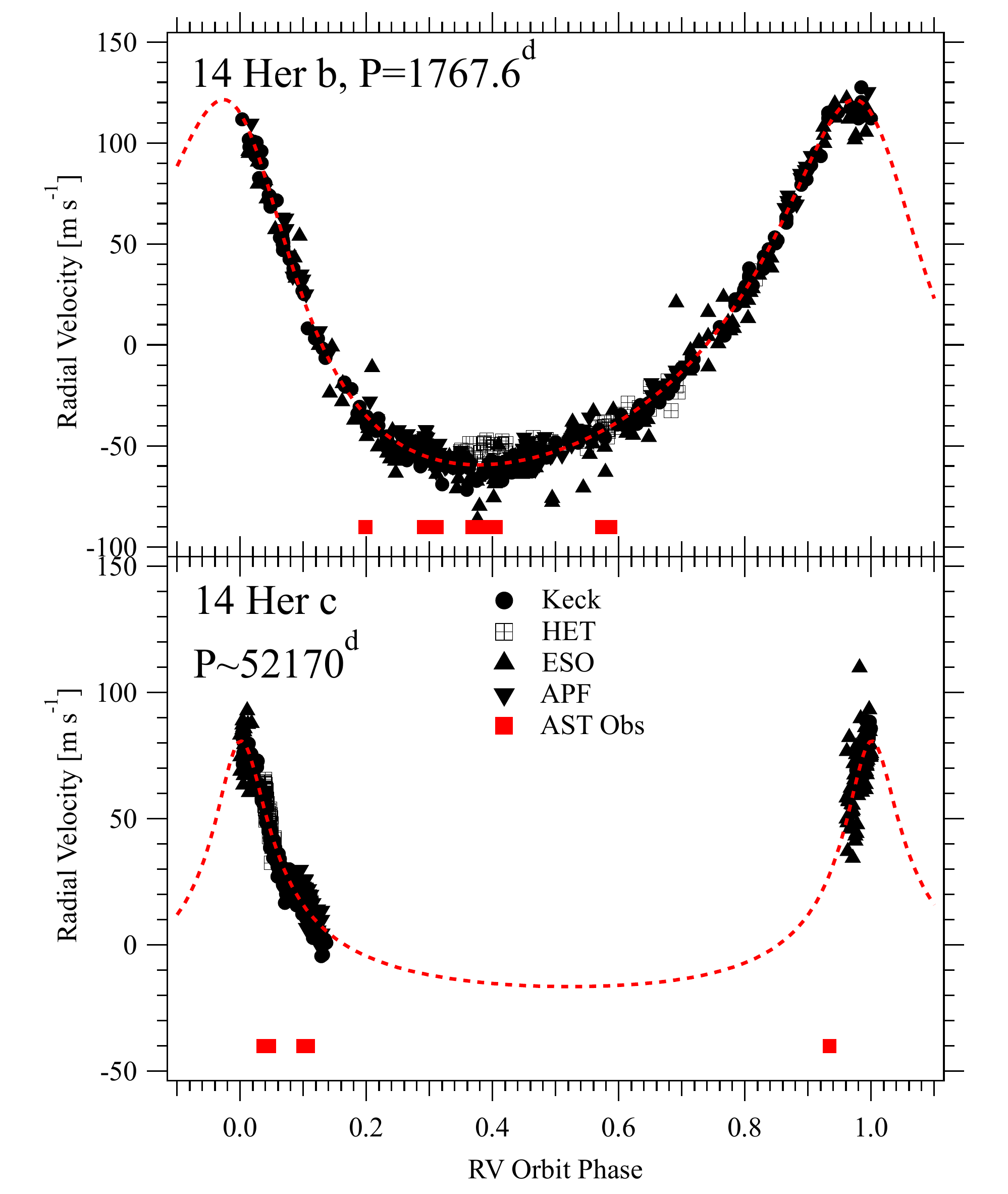}
\caption{RV measurements of \Hers from sources as indicated in the legend (and identified in Table ~\ref{tbl-RV}) phased to the orbital periods determined from  RV (Section~\ref{Ap1}). The dashed line is the RV predicted from the orbital parameters (Table~\ref{tbl-RVorb}). The red boxes denote corresponding phases of \HIP, FGS, DR1, and DR3 astrometry.}
\label{fig-RVbc}
\end{figure}
\end{center}

We  remove the \Her\,c RV orbit from the combined fit and phase the resulting RV (Figure~\ref{fig-RVbc}, top) to the \Her\,b orbital period. The bottom panel presents RV with component b subtracted, phased to the Table~\ref{tbl-RVorb}  \Her\,c period.  The red boxes in each panel indicate phases at which FGS astrometry took place, and for the \HIP,  \Gs DR1, and \Gs DR3 observations. 

\begin{center}
\begin{figure}
\includegraphics[width=6in]{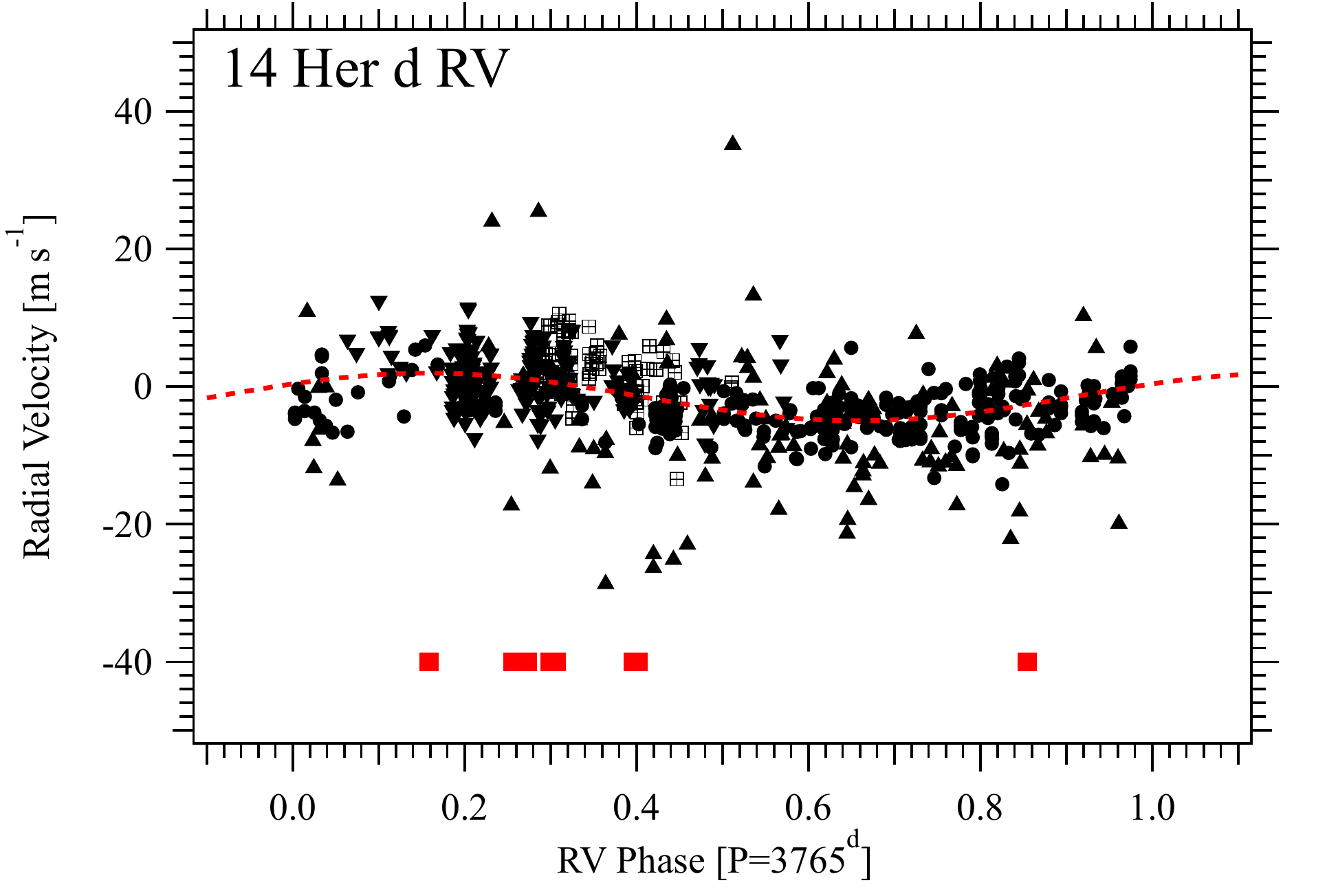}
\caption{ Candidate component d RV orbit (Table~\ref{tbl-HahD}) obtained from the RV$_{\rm d}$=RV$_{\rm all}$ - RV$_{\rm b}$ - RV$_{\rm c}$, phased to the Table~\ref{tbl-HahD} orbit period. A sine wave fit is plotted. RV source symbols from Figure~\ref{fig-RVbc}. The red boxes denote corresponding phases of \HIP, FGS, DR1, and DR3 astrometry.}
\label{fig-LSd}
\end{figure}
\end{center}

The residuals in Figure~\ref{fig-RVf} suggests a periodic pattern.  A
Lomb-Scargle periodogram of these residuals, after removing the RV contributions from \Her\,b,  and c, displays a very strong peak at P$\sim 3790^{\rm d}$.   We were able to determine the orbit-like parameters listed in Table~\ref{tbl-HahD} from these very noisy RV$_{\rm d}$ = RV$_{\rm all}$-RV$_{\rm b}$-RV$_{\rm c}$. We also list parameters for a simple sine wave fit 
\beq
RV=K*sin(((2\pi/P)*mJD)+\phi))
\eeq
to these same data. The two fits have similar $\chi^2$ values. Figure~\ref{fig-LSd} contains system RV with \Her\,b,c orbits subtracted, phased to $P=3765$, fit to the Table~\ref{tbl-HahD} sine wave.  
\cite{Hir21} identify a periodic signal at 3440$^{\rm d}$  that they interpret as a stellar activity cycle. To interpret the P$\sim 3790^{\rm d}$ signal as planetary in origin is premature.  See \cite{Ben10} and \cite{Hen13} to appreciate the value of continued RV monitoring in regards to the interpretation of low amplitude signals, and witness the demise of HD\,38529\,d.

\section{Astrometry}\label{AST}
\subsection{Astrometric Data}
For this study astrometric measurements came from Fine Guidance Sensor\,1r (\FGS\,1r), an upgraded \FGSs installed in 1997 during the second \HSTs servicing mission, from \Gs DR1\citep{Lin16}, DR3 \citep{Gai22}, and  from \HIP~\citep{Lee07a}. 
\subsubsection{\FGSs Data}
\begin{center}
\begin{figure}
\includegraphics[width=6in]{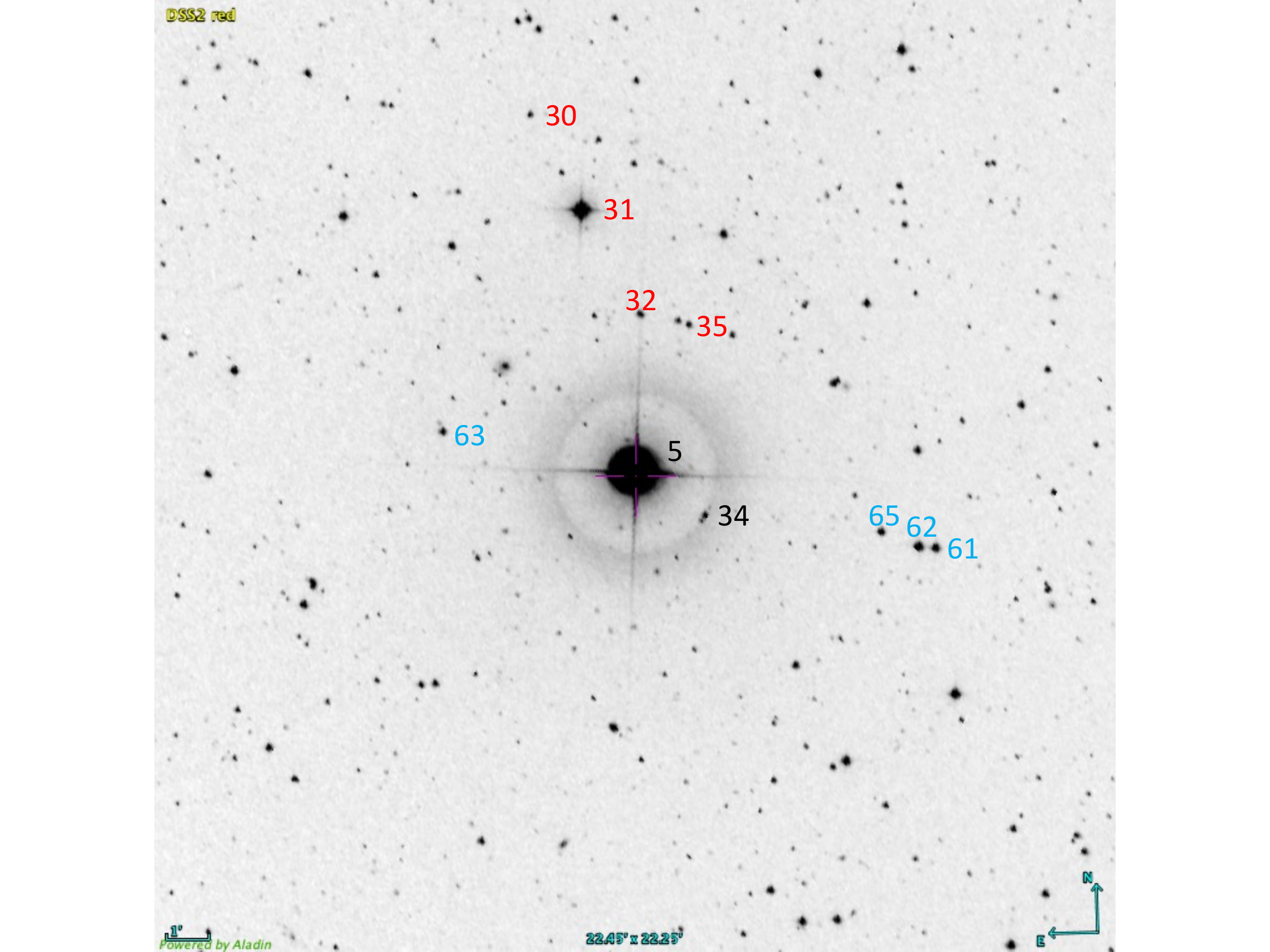}
\caption{Positions of \Hers (5) and astrometric reference stars. The FGS observed stars with \color{red}red ID numbers\color{black} ~in sets 1--6, and stars with \color{blue}blue IDs\color{black} ~in sets 7-17. Due to \HSTs gyro issues, \Hers and ref-34 were the only two stars observed in each set. The \Gs DR3 set 18 includes \Hers and all reference stars. }
\label{fig-Find}
\end{figure}
\end{center}

We utilized the fringe tracking mode (POS-mode; see Benedict et al. 2017 for a review of this technique)\nocite{Ben17} in this investigation. POS mode observations of a star 
have a typical duration of 60 seconds, during which over two thousand individual position measurements are collected. We estimate the astrometric centroid  by choosing the median measure, after filtering large outliers (caused by cosmic ray hits and particles trapped by the Earth's magnetic field). The standard deviation of the measures provides a measurement error. We refer to the aggregate of astrometric centroids of each star secured during one visibility period (typically on order 40 minutes) as a ``set". We identify the astrometric reference stars and science target in Figure~\ref{fig-Find}. 
Table~\ref{tbl-LOOF} lists mJD for all astrometry used in this investigation. These \FGSs data suffer from several deficiencies, which up until \G, have rendered them unable to produce the proposed \citep{Ben05p} result, that of a mass estimate for \Her\,b.  First, these data, collected from 2005.84 to 2007.19,  are significantly bunched, and provide for a companion period, $P>4.5$yr, effectively only three observational epochs; sets 1-6, sets 7-9, and sets 10-17.  Second, \HSTs gyro problems \cite[see][section 3, for details]{Ben10}~required that we switch astrometric reference frames during our sequence of scheduled observations. Sets 1-6 included \Hers and reference stars 30-35. Sets 7-17 included \Her, reference star 34, and reference stars 61- 65.  

\subsubsection{\Gs DR3}
To address both these problems, we incorporate three  epochs of \Hers astrometry from  \HIP, \Gs DR1, and DR3, and all reference star positions from \Gs DR3. To use them with the \FGSs data we derive standard coordinates, $\xi, \eta$, for reference stars and \Hers ICRS 2016.0 positions provided by DR3. Table~\ref{tbl-DR3p} contains those epoch 2016.0 positions. We also list the $RUWE$ (Renormalised Unit Weight Error) for each reference star. \cite{Sta21} find that the \Gs $RUWE$  robustly predicts unmodeled photocenter motion, even in the nominal "good" range of 1.0--1.4. 
Following \cite{vdK67}, 
\beq
\xi = {\cosr\delta\,\sinr\Delta\alpha \over (\sinr\delta \,\sinr\delta_0+\cosr\delta \,\cosr\delta_0\, \cosr\Delta\alpha)}
\eeq
\beq
\eta ={\sinr\delta\,\cosr\delta_0 - \cosr\delta\,\sinr\delta_0\,\cosr\Delta\alpha \over \sinr\delta\,\sinr\delta_0 + \cosr\delta\, \cosr\delta_0\,\cosr\Delta\alpha}
\eeq
where $\alpha, \delta$ are RA, Dec, $\alpha_0, \delta_0$ are the position of the  tangent point of the field (taken to be the average RA and Dec of the target and reference stars in Table~\ref{tbl-DR3p}), and $\Delta\alpha$ the RA distance from the tangent point for each star. The \FGSs astrometry pipeline produces standard coordinates in arc seconds. Hence, the DR3 $\xi, \eta$ in radians are  transformed to arc seconds.   
We derive the   \Hers \HIP~ and DR1 epoch by producing a difference (in arc seconds) between the DR3 ICRS position and the \HIP~and DR1 ICRS positions. 
The \Gs DR3 standard coordinates now constitute our master constraint plate. For past investigations we chose one of the \FGSs data sets as a master constraint plate. This did not work for our aggregate \Hers field data, given that the only stars in common to all \FGSs sets were  \Hers and reference star 34. The \Gs DR3 field contains all stars. We present DR3 reference star standard coordinates and a complete ensemble of \FGSs time-tagged  \Hers and reference star astrometric measurements, OFAD\footnote{The Optical Field Angle Distortion (OFAD) calibration {\citep{McA06}}.}- and intra-orbit drift-corrected, in Table~\ref{tbl-DATA}, along with calculated parallax factors in Right Ascension and Declination.  The \HIP~and DR1 \Hers positions, and the DR3 \Hers and reference star positions are ICRS barycentric, hence, require no parallax factors.  
 \subsubsection{\Gs DR1 and \HIP}
 For \Hers only we include ICRS positions from \Gs DR1 \citep{Lin16} and \HIP~\citep{Lee07a}. The latter extends the span of astrometry to 24.75 years. These measurements enter Table~\ref{tbl-DATA} as offsets from the DR3 ICRS position.

\subsection{Astrometry Modeling Priors}\label{CORR}
As in all of our previous \FGSs astrometry projects, e.g., \cite{Ben01,Ben07, Bea07, Mar10, McA10, Ben11, McA14, Ben16, Ben17b, Ben22b, Ben22c} we include as much prior information as possible in
our modeling.

We employ the following priors;
\begin{enumerate}

\item \textbf{Parallax:} 

This investigation  adopts DR3 values \citep{Gai22}. As in past investigations, we do not treat those values as being hardwired or absolute. Instead, we consider them to be quantities (Table~\ref{tbl-pimu}) introduced as observations with error. The average DR3 parallax error is 0.02 mas.  
Note that we utilize no parallax prior for \Her, an independent parallax having some value.
\item \textbf{Proper Motions:} For the reference stars we use the Table~\ref{tbl-pimu} proper motion priors from DR3 with median errors $\sigma_\mu \simeq 0.014$ mas. Simply relying on the DR3 values for \Hers might introduce a bias, given the limited DR3 time span and the potentially complicated perturbations from the known components.  We utilize no proper motion priors for \Her.

\item \textbf{Lateral Color Corrections:}  These corrections, entered into the model as data with errors, are identical to those used in \cite{Ben17b}. The $B-V$ values (Table~\ref{tbl-DR3p}) come from  measurements made with the New Mexico State University 1 m telescope \citep{Hol10}.

\item \textbf{Cross-Filter Corrections:}
{\it \FGS\,1r} contains a neutral density filter, reducing the brightness of \Hers by five magnitudes (from $V=6.6$ to $V=11.6$), permitting simultaneous modeling of \Hers  with far fainter reference stars 
 with $\langle V\rangle$ =12.8. These corrections, entered into the model as data with errors, are identical to those used in \cite{Ben17b}.
\end{enumerate}

\subsection{Modeling the \Hers Astrometric Reference Frame}  \label{AstRefs}
The astrometric reference frame for \Hers consists of nine stars (Table~\ref{tbl-DR3p}). 
The \Hers field (Figure~\ref{fig-Find}) exhibits the distribution of  astrometric reference stars (ref-30 through ref-65) used in this study. 
Due to \HSTs gyro difficulties,  \cite[see][section 3 for details]{Ben10}, the \Hers field was observed at a very limited range of spacecraft roll 
values (Table~\ref{tbl-DATA}). At each epoch we measured each available reference star 1--4 
times, and \Hers 3--5 times. Given the distribution of reference stars, FGS\,1r could observe only ref-34 and the science target, \Her,  at each epoch. The inclusion of a constraint plate derived from the  \Gs DR3 catalog permits the following astrometric analysis.

 Our choice of model (Equations 5-6) was driven entirely by the goodness of fit for the reference stars. We used no \Hers observations 
to determine the reference frame mapping coefficients, $A-F$.  We  solve  for 
roll, offsets, and independent scales along each axis.

\subsection{The  Model} \label{myMod}
From  
positional measurements we determine the scale, rotation, and offset ``plate
constants" relative to our adopted constraint epoch (\Gs DR3 ICRS 2016.0) for
each observation set. We employ GaussFit (Jefferys \etal 
1988) 
\nocite{Jef88} to minimize $\chi^2$. The solved equations of condition for the 
\Hers 
field are:
\begin{equation}
        x^\prime = x + lc_x(\it B-V) - \Delta XF_x
\end{equation}
\begin{equation}
        y^\prime = y + lc_y(\it B-V)  - \Delta XF_y
\end{equation}

\begin{equation}
\xi  = Ax^\prime + By' + C  - \mu_\alpha \Delta t  - P_\alpha\varpi - \sum_{n=1}^{2} O_{n,x} 
\end{equation}
\begin{equation}
\eta = Dx^\prime + Ey^\prime +  F  - \mu_\delta \Delta t  - P_\delta\varpi - \sum_{n=1}^{2} O_{n,y}
\end{equation}

\noindent 
Identifying terms, $\it x$ and $\it y$ are the measured coordinates from {\it HST};   $(B-V)$ is the Johnson $(B-V)$ color of each star; and $\it lc_x$ and $\it lc_y$ are the lateral color corrections, $\Delta XF_x$ and $\Delta XF_y$ are cross filter corrections applied only to \Her.  $A$, $B$, 
$D$, $E$,  
are scale and rotation plate constants, $C$ and $F$ are offsets; $\mu_\alpha$ and $\mu_\delta$ are proper motions; $\Delta t$ is the time difference from the constraint plate epoch; $P_\alpha$ and $P_\delta$ are parallax factors from a JPL Earth orbit predictor 
(Standish 1990)\nocite{Sta90}, version DE405;  and $\it \varpi$ is  the parallax.   $O_x$ and $O_y$, shown here for a two component planetary system,  are functions of the classic orbit parameters: $\alpha$, the perturbation semi major axis, $i$, inclination, $\epsilon$, eccentricity, $\omega$, argument of periastron, $\Omega$, longitude of ascending node, $P$, orbital period, and $T$, time of periastron passage for each included component  \citep{Hei78}. $\xi$ and $\eta$ are 
relative positions in RA and Dec that (once scale, rotation, parallax, the proper motions 
and the $O$ are determined) should not change with time.

\begin{figure}
\includegraphics[width=5in]{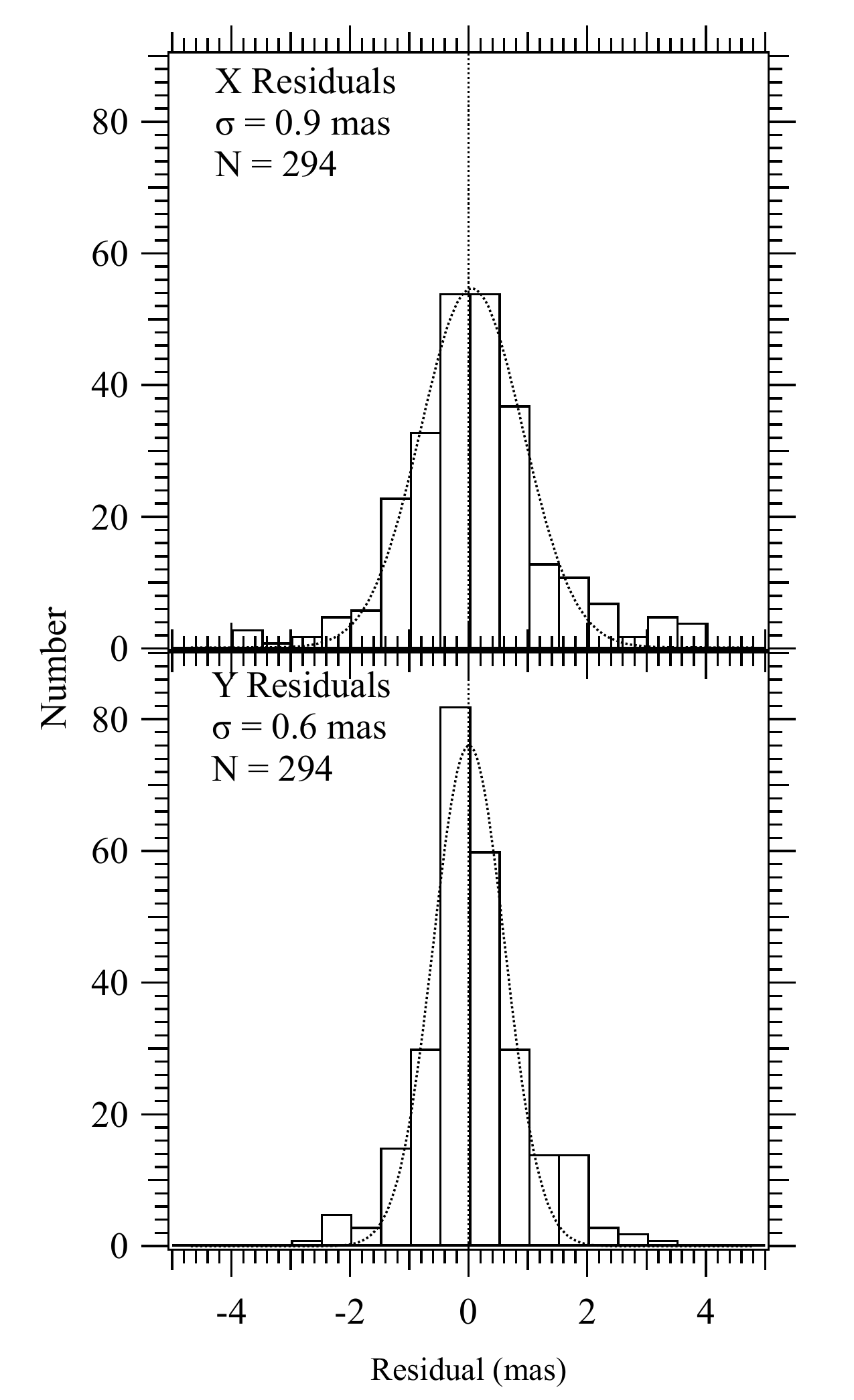}
\caption{Histograms of x and y residuals obtained by deriving the Equation 2-5 coefficients from  294 reference stars
measures (including the DR3 ICRS 2016.0 measurements),  while modeling reference star parallax and proper motion. The priors for this model had the published DR3 errors. Distributions are 
fit with gaussians with standard deviations, $\sigma$, indicated in each panel.} \label{fig-FGSH}
\end{figure}

   At this stage we model $only$ astrometry and $only$ the reference stars.
 From 
histograms of the  reference frame model 
astrometric residuals (Figure~\ref{fig-FGSH}) we conclude 
that we have a well-behaved reference frame solution exhibiting residuals with Gaussian distributions with dispersions $\sigma_{(\rm x,y)} = 0.9, 0.6$ mas. A  reference frame 
'catalog' from  \FGS\,1r and DR3 in $\xi$ and $\eta$ standard coordinates 
was determined with average uncertainties, $\langle\sigma_\xi \rangle= 0.07$ and $\langle\sigma_\eta \rangle = 0.07$ mas. Because  our constraint plate consists of \Gs DR3 RA, DEC, $\xi$ and $\eta$ are RA and DEC. 
Note that we removed ref-61 from the DR3 constraint plate due to its very high $RUWE$ value (Table~\ref{tbl-DR3p}). The behavior of the ref-61 FGS\,1r residuals indicate nothing  systematic, other than having the largest final positional uncertainties, $\sigma_\xi = 0.57$ and $\sigma_\eta = 0.53$ mas. Lastly, we obtained a final reference frame mapping by removing ref-30. This resulted in  a $55 \%$ reduction in $\chi^2$, compared to including that reference star.

 At this step in the  analysis the astrometry knows nothing of the RV detections (Table~\ref{tbl-RVorb}). With our derived  $A$, $B$, 
 $D$, $E$, 
 $C$, and $F$ 
 we transform the \Hers astrometric measurements, applying $A$ through $F$ as constants, solving only for \Hers proper motion and parallax, using no priors for \Her.  Table~\ref{tbl-SUM}  compares values for the parallax and proper motion of \Her 
~from {\it HST} and \G~\citep{Gai22}. We note a  significant disagreement in the  proper motion vector ($\vec{\mu}$) absolute magnitude and the parallax values. This could be explained both by our non-global proper motion measured against a small sample of reference stars, and the limited duration of both astrometric studies, possibly affected by the companion perturbations.
Alternatively, the mismatch between our proper motion, established through measurements taken from 1991.25 (\HIP), 2005.84 to 2007.19 (\FGS) and the \Gs DR1, DR3 values, a result of a campaign spanning 2014.6 - 2017.4, could indicate acceleration due to the companions, as might the $RUWE=1.82$.  Furthermore, \cite{Bra21} finds a very high $\chi^2>1000$,  when solving a model assuming no proper motion change, comparing \HIP~ with \G, indicating significant \Hers acceleration over a roughly 25 year time span. 
\section{Perturbation Orbits and Mass Estimates}\label{ORBIT}
{ \subsection{\Her\,{\lowercase{b,c}}}}
We next employ Equations 7-8, looking  for astrometric evidence of \Her\,b and c. With our derived  $A$, $B$, 
 $D$, $E$, 
 $C$, and $F$ 
 we transform the \Hers astrometric measurements, applying $A$ through $F$ as constants, now solving for \Hers proper motion, parallax, and b, c orbits, again, using no parallax and proper motion priors for \Her.
We force astrometry and RV to describe the same perturbation
through a constraint \citep[e.g.][]{Pou00} for a perturbing companion. 
\begin{equation}
\displaystyle{{\alpha_{\rm b}~sin\,i_{\rm b} \over \varpi_{abs}} = {P_{\rm b} K_{\rm b} (1 -
e_{\rm b}^2)^{1/2}\over2\pi\times4.7405}} 
\label{PJ}
\end{equation}
\noindent Using  \Her\,b as an example, Equation~\ref{PJ} contains quantities derived from astrometry (parallax, $\varpi_{abs}$, 
host star perturbation orbit size, $\alpha$, and inclination, $i$) on the left hand side (LHS), and 
quantities  derivable from both (the period, $P$ and eccentricity, $\epsilon$), or only radial 
velocities  (the RV amplitude of the primary, $K$, induced by a companion), on the right hand side (RHS).

The RV provide far higher cadence coverage of  the \Her\,b and c  perturbations than do the far sparser astrometric measurements, providing essential
support for determining $P$, $\epsilon$, $K$, $\omega$, and $T$.
For these orbit determinations, we introduce the Table~\ref{tbl-RVorb} RV-provided values as observations with error. We use the  Equation~\ref{PJ} relationship between the astrometry and the RV, but hold no orbital or astrometric parameters as constants. Our solutions do not converge unless there is a measurable signal.
 Table~\ref{ASTRVorb} contains the resulting orbits. For \Her\,b the astrometry has improved the precision of RV-derived orbital parameters,
 significantly reducing the $1-\sigma$ errors on $P_{\rm b}$, $\epsilon_{\rm b}$, $K_{\rm b}$, $\omega_{\rm b}$, and $T_{\rm b}$, listed in Table~\ref{tbl-RVorb}. For \Her\,c we find improvements in all but $\epsilon_{\rm c}$ and $\omega_{\rm c}$.

\begin{figure}
\includegraphics[width=5in]{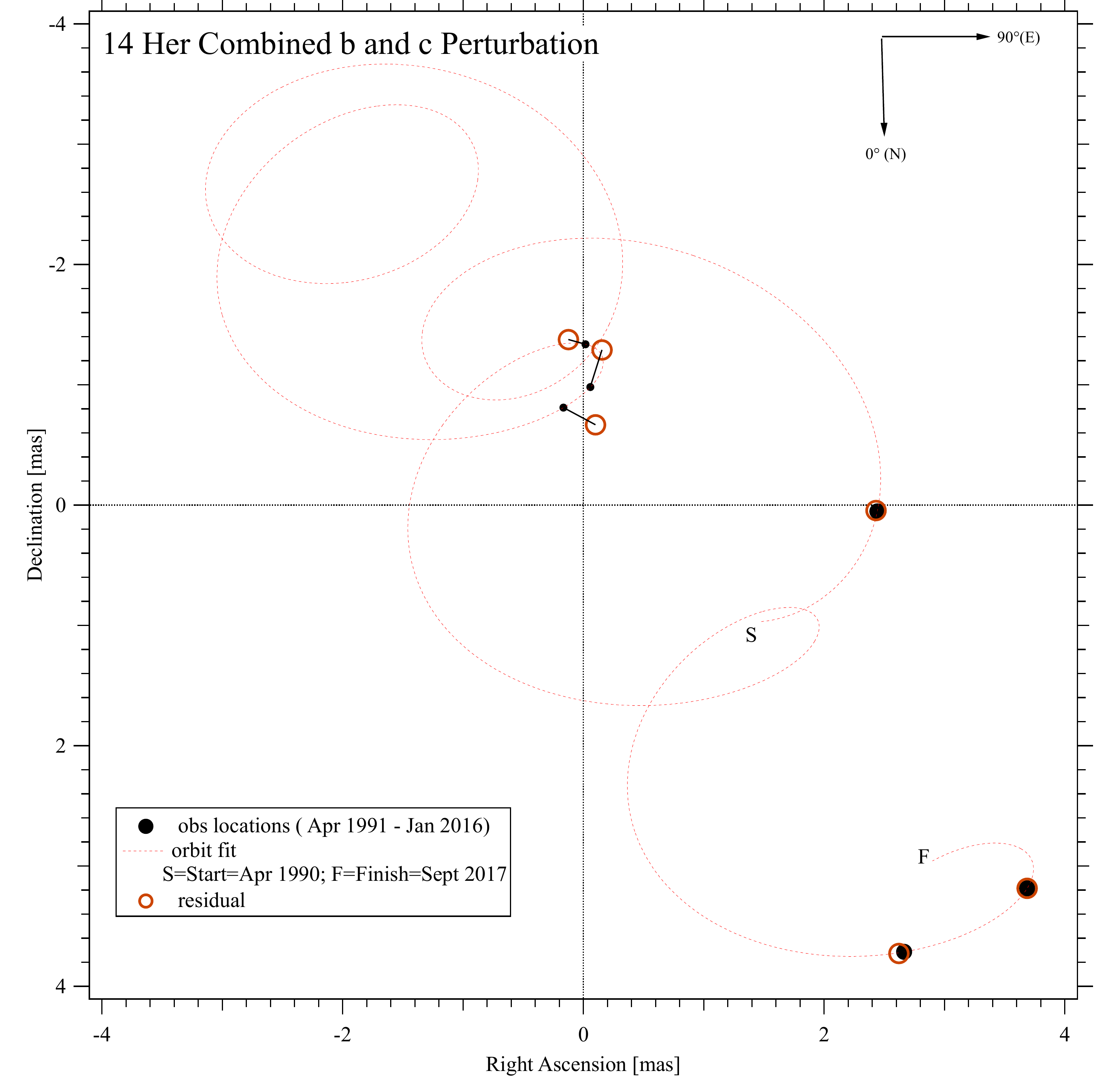}
\caption{ Residuals to the \Her\,b and c perturbation described by the Table~\ref{ASTRVorb} final orbital elements. Normal points (\Large \color{red}$\circ$\color{black}\normalsize), as described in Section~\ref{ORBIT},    are near their calculated locations (FGS $\bullet$; HIPP, DR1, DR2 {\Large$\bullet$\normalsize}) on the  orbit. The residual rms is 0.15 mas in R.A. and 0.13 mas in Dec.   \label{fig-ASTorb}}
\end{figure}

The mutual inclination, $\Phi$, of the b and c orbits can be determined through \citep{Kop59, Mut06}
\beq
cos\,\Phi = cos\,i_b cos\,i_c + sin\,i_b sin\,i_c cos(\Omega_b - \Omega_c) \label{MI}
\eeq
\noindent where $i_b$ and $i_c$ are the orbital inclinations and $\Omega_b$ and  $\Omega_c$  are the longitudes of their ascending nodes.
Our modeling yields a significant lack of coplanarity with $\Phi = 62 \pm 12 \arcdeg$. 
Figure~\ref{fig-ASTorb} shows the perturbation due to both companions, the epochs of observation and the residuals. We have collapsed the 
final 106 residuals from 17 \FGSs epochs, first by averaging the 17 epochs (averaging 5 residuals per epoch) down to 17 residuals, We then averaged  the epochs 1-6, epochs 7-9, and epochs  10-17 residuals to obtain the plotted values. The (effectively) three \FGSs epochs provide a sub-standard observation set for orbit determination. The 1991.25 \HIP, 2015.0 DR1, and 2016.0 DR3 epochs improve the parameter accuracy and errors, providing, combined with the \FGSs epochs,  over 5 wraps of the orbit for $P_{\rm b}$. 

To determine a \Her\,b mass we find an \m$_{\rm b}$ which satisfies this relation \citep{Hei78}
\begin{eqnarray}
 f({\rm \cal{M}}) = ({\rm \cal{M}}_{\rm b} \sin{i})^3 / ({\rm \cal{M}}_* + {\rm \cal{M}}_{\rm b}) ^2 \\
 R = 1.036\times10^{-7}K_{\rm p}P(1-\epsilon^2)^{3/2}\\
 f({M})=R
 \label{mf}
\end{eqnarray} 
The planetary mass depends on the mass of the primary star. Assuming \m$_*=0.98\pm0.04$\msun \citep{Bar21}, our orbit yields \m$_{\rm b}= 8.5^{+1.0}_{-0.8}$\mjupe. A similar analysis for \Her\,c yields \m$_{\rm c}= 7.1^{+1.0}_{-0.6}$\mjupe.

Modeling only the \Her\,b perturbation while including DR3 parallax and proper motion priors for \Hers increases the $\chi^2/DOF$ by a factor of 3.7, yielding similar
perturbation parameters, but with significantly increased parameter errors. Hence, for \Her\,b,c we adopt orbit results without DR3 priors for \Her.

 \subsection{\Her\,{\lowercase{d}}}
Candidate \Her\,d, with P$\sim 3765^{\rm d}$, producing an RV amplitude of $K_{\rm d}=3.8$ \ms, would have an \msini$\sim0.28$\mjup minimum mass, which for a circular orbit would produce a perturbation, $\alpha_{\rm d}=70$ microarcseconds, undetectable with \FGSs data. Attempts to solve for an astrometric signature (perturbation orbit size, $\alpha$, inclination, $i$, $\Omega$, longitude of ascending node) for \Her\,d were unsuccessful. 

Many stars produce long-period cycles of activity \citep[e.g.][]{Bal95}, and these cycles can affect RV measurements \citep{Cos21}. Longer-period activity cycles are more likely for longer stellar rotation periods \citep{Ola16}. Does sufficient evidence exist to unambiguously ascribe the P$\sim 3765^{\rm d}$ RV signal to a stellar activity cycle?  \cite{Hir21}, working with an activity parameter derived from the strengths of the Calcium H and K lines, $S_{HK}$, find a P$\sim 3440^{\rm d}$, a variation directly attributable to  stellar activity. For \Her, $\langle S_{HK}\rangle=0.151$ \citep{Dunc91}. The Sun has  $\langle S_{HK}\rangle=0.170$ \citep{Hal09,Ege17}. \cite{Bar21} list another activity parameter for \Her, $logR'_{HK}=-4.94$ \citep{Mor19}. For the Sun $logR'_{HK}=-4.96$ \citep{Hal09}. The rotation periods for \Hers and the Sun are $P_{rot}=29.5^{\rm d}$ and $P_{rot}=27.3^{\rm d}$ with similar estimated ages of 4.6Gyr \citep{Bar21, Bon02}. The Sun has a sunspot cycle with an average period, $P\sim11$yr, with a range 8-17 yr \citep{Hat15,Uso21}. Given the activity level agreement between the Sun and \Her, the \Hers RV signal for candidate d ($P_{\rm d}=10.4$yr) seems more likely activity than a perturbation. 

We hypothesize that the activity period disagreement between \cite{Hir21} and that from Section~\ref{Ap1} might come from our access to additional HET RV data. We also included earlier, less precise ELODIE RV measurements, which extended the RV time span by over 9 years.

\section{Discussion}\label{Disc}
The Table~\ref{ASTRVorb} astrometry-derived parameters, $\alpha$, $i$, and $\Omega$ values for the \Her\,b,c perturbations differ  from \cite{Fen22}, but agree more closely with  \cite{Bar21}. The MCMC sampling approach used by \cite{Bar21} also finds a significant mutual inclination, $\Phi = 96_{-37}^{+29} \arcdeg$, in general agreement with our value, $\Phi = 62 \pm 12 \arcdeg$. 
Using Equation~\ref{MI} and \cite{Fen22} values yields a smaller mutual inclination, $\Phi = 20\pm2\arcdeg$. 

Our \m$_{\rm b}$ and \m$_{\rm c}$ values (Table~\ref{ASTRVorb}) agree with \cite{Bar21} within their respective errors, and agree for \m$_{\rm b}$ with \cite{Fen22}. Regarding \m$_{\rm c}$, both our and that of \cite{Bar21}  disagree with the \cite{Fen22} value, \m$_{\rm c}=5.0^{+1.1}_{-0.9}$, understandable given the \cite{Fen22} far shorter period for \Her\,c, $P_{\rm c}=15732^{+1896}_{-2654}$ d.

\cite{Ben17} reviews \HSTs \FGSs past astrometry to characterize the perturbations due to candidate planets. More recent results include masses for HD202206 B and c, a circumbinary brown dwarf system \citep{Ben17b}, and null results for companions to $\mu$ Ara of any mass \citep{Ben22c}. During the publication process for the latter paper, a referee pointed out the possibility that inclinations obtained  with \FGSs astrometry were biased towards low values. We explored this at length in \cite{Ben22c}, but 
could not identify any flaws in our modeling. Our  agreement with \cite{Bar21} for \Her\,b,c provides additional support for the soundness of our analyses. 

However, to further explore this possible bias, we  compare, via a Kolmogorov–Smirnov (K-S) test, a cumulative distribution function (CDF) for 177 companion inclinations in \cite{Fen22} to  our \FGS-derived inclinations, now including the \Her\,b,c results.  To produce the CDF we put all inclinations on a 0-90$\arcdeg$ scale by applying an offset to  inclinations over 90$\arcdeg$; $i_{\rm corr}=90\arcdeg-{\rm mod}(i,90\arcdeg)$. A K-S test  produces a test statistic, D, a critical value, C, and a p value, PV. Values of D less than C support the null hypothesis, as does a p value larger than an adopted significance level, $\alpha=0.05$.  
The K-S test of \FGSs against \cite{Fen22} inclinations suggest that both samples were drawn from the same parent distribution, D less than C with PV$>\alpha$ (Table~\ref{tbl-KSres}). 

\begin{figure}
\includegraphics[width=5in]{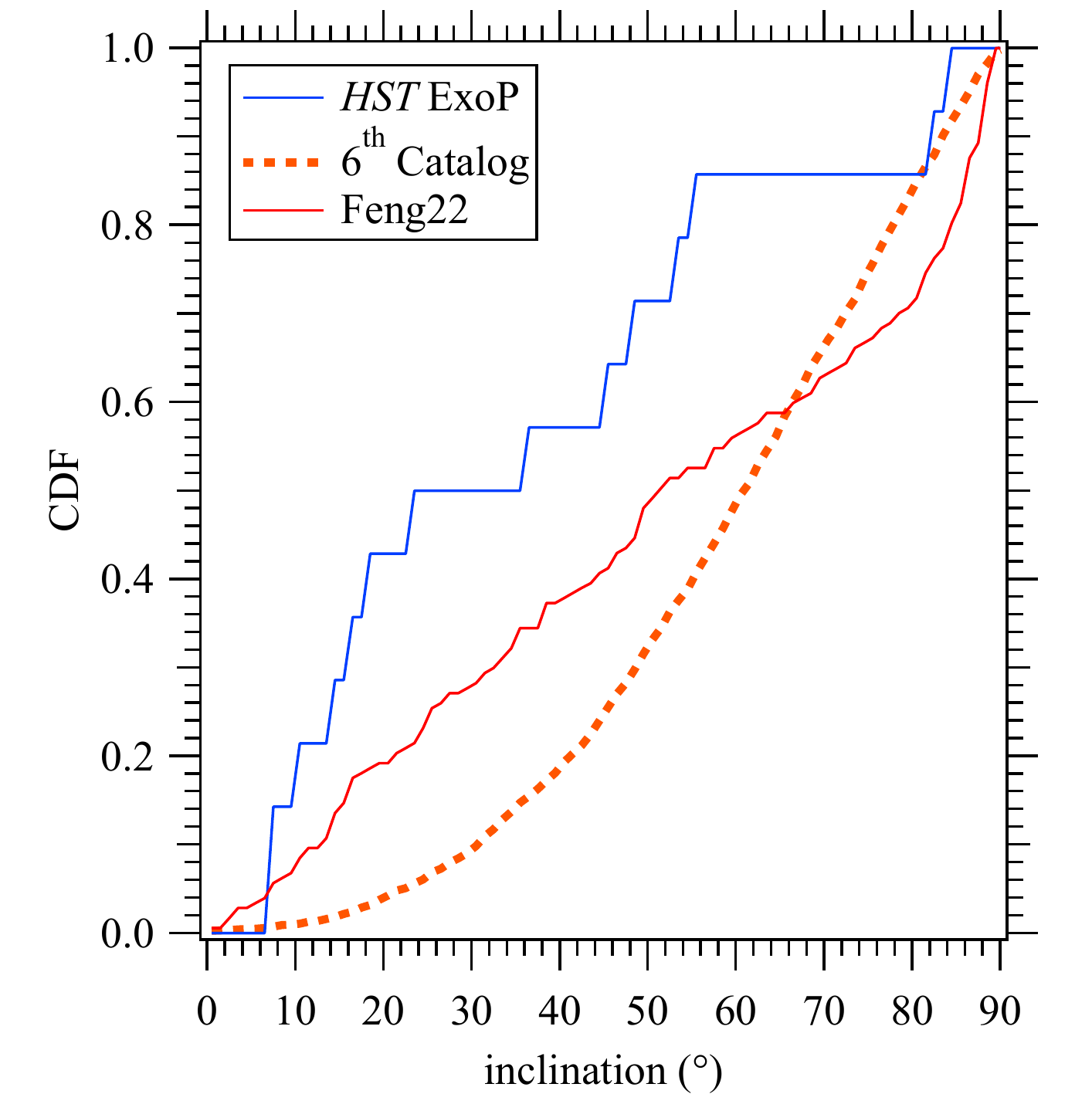}
\caption{CDFs for; entire inclination set from the 6$^{\rm th}$ Visual Binary Star Catalog  \citep{Har01};  exoplanet perturbation inclinations   \cite[][(table 11)]{Ben22c}, now including \Her\,b,c; and exoplanet perturbation inclinations from \cite{Fen22}. KS test results indicate  that  neither our exoplanet inclination distributions nor the \cite{Fen22} inclination distributions are drawn from the same parent population as the 6$^{\rm th}$ Catalog binary inclination population, presumed to be random.   \label{fig-KS}}
\end{figure}

We then compare the \cite{Fen22} inclinations with a set of 3214 binary star inclinations, presumed to be purely random in distribution, \citep[6th Catalog of Visual Binary Stars,][]{Har01}.
We find  the null hypothesis that  \cite{Fen22} inclinations are as random as the 6$^{\rm th}$ Catalog inclinations is not supported, D  marginally larger than C, and the PV far lower than the significance level, $\alpha$ (Table~\ref{tbl-KSres}). Similarly retesting the \FGS-derived inclinations continues to indicate a bias to lower inclinations. Both the \FGSs and  the \cite{Fen22} inclinations seem  inconsistent with a sample with a random distribution of inclinations. We show these CDF  in Figure~\ref{fig-KS}. These K-S tests suggest either that observed exoplanetary systems, first discovered via RV,  have astrometrically measured inclinations with a real  bias towards smaller values, or that both  \cite{Fen22} and the \FGSs  have unresolved issues with their analyses. See \cite{Pou01} for one possible explanation and \cite[][section 5.2.2]{Ben22b} for a counter argument in support of the  \FGS.

\section{Summary} \label{Summ}

For  the \Hers system from  modeling 25 years of ground-based RV and  the results from models which utilizes 17 epochs of {\it HST/FGS}, one epoch each of \Gs DR1 and DR3, and one epoch of \HIP~ astrometry all spanning 25.75 years, we find;
\begin{enumerate}
\item derived from only the augmented body of RV data now including HET measurements,  improved companion orbital elements ($P$, $\epsilon$, $\omega$, $T_0$, $K$),

\item after subtracting the \Her\,b and c RV signals, evidence for a signal with $P\sim3765^{\rm d}$,  tentatively identified as an activity cycle,

\item from a model 
containing no proper motion, no parallax priors, nor  perturbing \Her\,b,c orbits  for \Hers a parallax, $\pi_{abs}=54.14\pm0.22$ mas, disagreeing with the \Gs DR3 value, and a  proper motion relative to a \Gs DR3 reference frame, $\vec{\mu} = 325.90$ mas  yr$^{-1}$
with a position angle, P.A. = $155\fdg95$, differing by +0.97 mas yr$^{-1}$ and $-0\fdg13$ compared to \Gs DR3,

\item from a model 
containing neither \Hers proper motion nor parallax priors but with   perturbing \Her\,b, c orbits  for \Her, a parallax, $\pi_{abs}=54.77\pm0.22$ mas,  and a  proper motion relative to a \Gs DR3 reference frame, $\vec{\mu} = 326.11$ mas  yr$^{-1}$
with a position angle, P.A. = $156\fdg04$, differing by -1.2 mas yr$^{-1}$ and $+0\fdg04$ compared to \Gs DR3,

\item that model, using the RV-derived $P_{\rm b,c}$, $\epsilon_{\rm b,c}$, $K_{\rm b,c}$, $\omega_{\rm b,c}$, and $T_{\rm b,c}$ orbital parameters as observations with error, yields $\alpha_{\rm b}=1.3\pm0.1$ mas, $i_{\rm b}=35\fdg7\pm3\fdg2$,  $\Omega_{\rm b}=276\arcdeg \pm 5\arcdeg$, and $\alpha_{\rm c}=10.3\pm0.7$ mas, $i_{\rm c}=82\pm14\arcdeg$,  $\Omega_{\rm c}=224\arcdeg \pm 9\arcdeg$ with errors on $P$, $\epsilon$, $K$, $\omega$, and $T$ smaller than those obtained only from RV,

\item a mass for \Her\,b, \m$_{\rm b}=8.5^{+1.0}_{-0.8}$\mjup, and for \Her\,c, \m$_{\rm c}=7.1^{+1.0}_{-0.6}$\mjup

\item a significant lack of coplanarity with $\Phi = 62 \pm 12 \arcdeg$

\end{enumerate}

A  combination of \FGSs and RV data with future \Gs data releases can  produce significantly improved companion orbits and masses for  \Her\, b and c, and confirm a stellar activity source for the $P\sim 3780$ d RV signal. 

All exoplanetary systems investigated through \HSTs\FGSs astrometry were first identified as possible targets via RV, as were the systems investigated in \cite{Fen22}. We find evidence that these two independent investigations of exoplanet host stars have companions that exhibit a bias towards lower inclinations. Perhaps a large number of systems discovered by \G, without previous RV
evidence for their existence, can assist in identifying a cause for this bias.

\begin{acknowledgments}

We thank an anonymous referee for their suggestions. This work is based on observations made with the NASA/ESA Hubble Space Telescope, obtained at the Space Telescope Science Institute. Support for this work was provided by NASA through grants 11210 and 11788
 from the Space Telescope 
Science Institute, which is operated
by the Association of Universities for Research in Astronomy, Inc., under
NASA contract NAS5-26555.    This research has made use of the {\it SIMBAD} and {\it Vizier} databases, 
operated at Centre Donnees Stellaires, Strasbourg, France; Aladin, developed and maintained at CDS;  and NASA's truly essential Astrophysics Data System Abstract Service. This work has made use of data from the European Space Agency (ESA)
mission {\it Gaia} (\url{http://www.cosmos.esa.int/gaia}), processed by
the {\it Gaia} Data Processing and Analysis Consortium (DPAC,
\url{http://www.cosmos.esa.int/web/gaia/dpac/consortium}). Funding
for the DPAC has been provided by national institutions, in particular
the institutions participating in the {\it Gaia} Multilateral Agreement. GFB thanks Dr. Paul Butler for cheerfully offered observation accounting assistance. Thanks to Dr. Tom Harrison for securing the  reference star $B-V$ photometry.
\end{acknowledgments}

\bibliography{/Users/gfbenedict/Active/myMaster}

\clearpage
\clearpage

\begin{deluxetable}{l r c c c}
\tablecaption{The RV Data Sets \label{tbl-RV}}
\tablewidth{0in}
\tablehead{ 
\colhead{Data Set}& 
\colhead{Coverage}& 
\colhead{Nobs} & 
\colhead{RMS [\ms]} &
\colhead{RV ZP [\ms] \tablenotemark{a}}
}
\startdata
\hline
ESO&1995.09-2003.59&117&9.9&72.3$\pm$1.2\\
Keck  &2004.64-2020.15 &  231   & 3.3 &0 \\
HET  &2006.04-2008.31 &77  &4.9 &5.4$\pm$0.6\\
APF  &2014.01-2019.28 &  208  &3.8  &-18.4$\pm$0.2\\
\hline
&total&684& &
\enddata 
\tablenotetext{a}{RV zero-point adjustment values relative to Keck.}
\end{deluxetable}
\clearpage
\begin{deluxetable}{c c c c}
\tablewidth{2.75in}
\tablecaption{ RV Data\tablenotemark{a}  \label{tbl-ALLRV}}
\tablehead{ 
\colhead{mJD}& 
\colhead{RV (m/s)}& 
\colhead{$\pm$ error}\tablenotemark{b} &
\colhead{source}\tablenotemark{c}
} 
\startdata
53752.5331&3.9&8.9& HET\\
53753.5069&3.2&9.0& HET\\
53754.4962&4.8&9.2& HET\\
53755.5097&1.9&9.0& HET\\
53756.5094&0.1&9.0& HET\\
53764.4816&2.3&8.9& HET\\
53766.4967&1.8&8.9& HET\\
53768.4681&3.5&8.8& HET\\
53778.4754&7.3&8.8& HET\\
53780.4533&-2.3&8.8& HET\\
53787.4326&7.1&10.1& HET\\
53794.4119&0.1&8.9& HET\\
53800.3741&-2.2&8.8& HET\\
53808.3481&2.4&9.3& HET
\enddata
\tablenotetext{a}{Full table available on-line.}
\tablenotetext{b}{Errors adjusted to achieve
$\chi^2/DOF$ near unity.}
\tablenotetext{c}{HET = Hobby-Eberly Telescope, ESO = ELODIE, Keck = HIRES, APF = Automatic Planet Finder}
\end{deluxetable}
\clearpage
\begin{deluxetable}{l r r}
\tablewidth{0in}
\tablecaption{14 Her\,b,c Orbital Parameters from RV \label{tbl-RVorb}}
\tablehead{
\colhead{Parameter} & 
\colhead{b} & 
\colhead{c} 
}
\startdata
$P$ [d]& 1767.8$\pm$ 0.6&52172 $\pm$  2326\\
$P$ [yr]& 4.840$\pm$ 0.002&143 $\pm$  6\\
$T$ [d] &51368 $\pm$ 2 &51781 $\pm$ 73 \\
$e$  & 0.372$\pm$ 0.003 & 0.63 $\pm$ 0.01 \\
$\omega$ (\arcdeg) &22.3$\pm$ 0.4 &-2$\pm$2  \\ 
$K$ (\ms) & 90.3 $\pm$ 0.4 & 51 $\pm$ 1 
\enddata
\end{deluxetable}
\clearpage
\begin{deluxetable}{l r r}
\tablewidth{0in}
\tablecaption{14 Her {\lowercase{d}} Orbital and Sin Wave Parameters from RV \label{tbl-HahD}}
\tablehead{
\colhead{Parameter} & 
\colhead{orbit} & 
\colhead{Sin Wave} 
}
\startdata
$P$ [d]& 3765$\pm$ 77&3789 $\pm$  71\\
$T$ [d] &56425 $\pm$ 183 &- \\
$e$  & 0.17$\pm$ 0.06 & - \\
$\omega$\tablenotemark{a} (\arcdeg) &-47$\pm$ 18 &52$\pm$2  \\ 
$K$ (\ms) & 3.8 $\pm$ 0.2 & 3.5 $\pm$ 0.2 
\enddata
\tablenotetext{a}{$\phi$ in Equation 1}
\end{deluxetable}
\clearpage
\begin{center}
\begin{deluxetable}{r l}
\tablewidth{2in}
\tablecaption{Log of \Her ~Astrometry Observations\label{tbl-LOOF}}
\tablehead{
\colhead{Set}&
\colhead{mJD}
}
\startdata
1&53681.092\\
2&53682.954\\
3&53683.886\\
4&53684.885\\
5&53685.883\\
6&53686.881\\
7&53786.216\\
8&53804.872\\
9&53813.938\\
10&54150.063\\
11&54150.992\\
12&54152.060\\
13&54157.718\\
14&54163.045\\
15&54168.705\\
16&54171.102\\
17&54172.633\\
18\tablenotemark{a}&48347.313\\
18\tablenotemark{b}&57023.000\\
18\tablenotemark{c}&57388.000
\enddata
\tablenotetext{a}{ICRS epoch 1991.25 from \HIP}
\tablenotetext{b}{ICRS Epoch 2015.0 from \Gs DR1}
\tablenotetext{c}{ICRS Epoch 2016.0 from \Gs DR3}
\end{deluxetable}
\end{center}
\clearpage
\begin{deluxetable}{r l l l l l l l}
\tablewidth{0in}
\tablecaption{  Star Positions\tablenotemark{a} from \Gs DR3  \label{tbl-DR3p}}
\tablehead{
\colhead{ID}&
\colhead{RA [$\arcdeg$]}&
\colhead{RA err [mas]}&
\colhead{Dec[$\arcdeg$}&
\colhead{Dec err [mas]}&
\colhead{G [mag]\tablenotemark{b}}&
\colhead{RUWE\tablenotemark{c}} &
\colhead{$B-V$\tablenotemark{d}}
}
\startdata
\Her&242.6021268&0.02&43.8163208&0.03&6.40&1.819&0.87\\
30\tablenotemark{e}&242.6577288&0.01&43.9581798&0.01&14.15&1.016&0.99\\
31&242.6305334&0.01&43.9208867&0.01&9.43&1.056&0.52\\
32&242.5982813&0.01&43.8801370&0.01&13.64&1.061&0.71\\
34&242.5643202&0.01&43.8021963&0.01&14.20&0.974&0.52\\
35&242.5725899&0.01&43.8758601&0.02&14.55&0.963&0.87\\
61\tablenotemark{f}&242.4391655&0.15&43.7893724&0.17&12.94&14.457&0.87\\
62&242.4481893&0.01&43.7898342&0.01&12.29&1.034&0.70\\
63&242.7050172&0.01&43.8342381&0.01&13.83&1.013&0.87\\
65&242.4685474&0.01&43.7949531&0.01&13.60&1.029&1.17
\enddata
\tablenotetext{a}{Epoch 2016.0 ICRS Positions from DR3}
\tablenotetext{b}{G Magnitude from DR3; all errors 0.0028 mag.}
\tablenotetext{c}{Reduced Unit Weight Error.}
\tablenotetext{d}{all errors 0.03 mag.}
 \tablenotetext{e}{Not included in modeling due to high \FGSs  residuals.}
\tablenotetext{f}{Not included in DR3 constraint plate due to high $RUWE$.}
\end{deluxetable}
\clearpage
\begin{deluxetable}{l l l r r r r r r r r}
\tabletypesize{\tiny}
\tablewidth{0in}
\tablecaption{\Hers Field Astrometry\tablenotemark{a} \label{tbl-DATA}}
\tablehead{ 
\colhead{Set}& 
\colhead{Star}& 
\colhead{ID} &
\colhead{field roll\tablenotemark{b}} &
\colhead{X} &
\colhead{Y} &
\colhead{$\sigma_X$} &
\colhead{$\sigma_Y$} &
\colhead{t$_{\rm obs}$} &
\colhead{P$_{\alpha}$} &
\colhead{P$_{\delta}$}
}
\startdata
18&\Her&DR3&0&86.96035&-107.53965&2.22E-05&2.68E-05&57388.0&0&0\\
18&\Her&DR1&0&86.82854&-107.24292&2.12E-04&2.23E-04&57023.0&0&0\\
18&\Her&HIPP&0&83.678864&-100.16482&2.7E-04&3.0E-04&48347.3&0&0\\
18&30&DR3&0&230.84356&403.26011&1.12E-05&1.18E-05&57388.0&0&0\\
18&31&DR3&0&160.46832&268.94001&1.02E-05&1.10E-05&57388.0&0&0\\
18&32&DR3&0&76.88882&122.19479&1.01E-05&1.11E-05&57388.0&0&0\\
18&34&DR3&0&-11.24992&-158.40539&1.22E-05&1.31E-05&57388.0&0&0\\
18&35&DR3&0&10.22438&106.78428&1.37E-05&1.52E-05&57388.0&0&0\\
18&62&DR3&0&-313.05087&-202.68143&9.10E-06&9.90E-06&57388.0&0&0\\
18&63&DR3&0&354.12504&-42.76332&1.04E-05&1.13E-05&57388.0&0&0\\
18&65&DR3&0&-260.12227&-184.32370&1.00E-05&1.13E-05&57388.0&0&0\\
1&5&F9D23701M\tablenotemark{c}&159.836&-89.59243&-90.29798&2.22E-03&2.43E-03&53681.07782&-0.463950&-0.820011\\
1&5&F9D23707M&159.836&-89.59375&-90.29817&2.01E-03&2.32E-03&53681.0858&-0.463782&-0.820035\\
1&5&F9D2370DM&159.836&-89.59390&-90.29833&1.96E-03&2.25E-03&53681.0936&-0.463642&-0.820063\\
1&5&F9D2370HM&159.836&-89.59469&-90.29905&1.94E-03&2.41E-03&53681.09898&-0.463550&-0.820090\\
1&5&F9D2370LM&159.836&-89.59578&-90.29930&2.02E-03&2.35E-03&53681.10392&-0.463462&-0.820124\\
1&30&F9D23706M&159.836&437.31868&-52.27713&3.08E-03&2.53E-03&53681.08441&-0.466121&-0.820927\\
1&30&F9D2370CM&159.836&437.32002&-52.27917&2.96E-03&2.53E-03&53681.09221&-0.465977&-0.820953\\
1&30&F9D2370GM&159.836&437.31986&-52.27674&2.96E-03&2.57E-03&53681.09758&-0.465885&-0.820978\\
1&31&F9D23705M&159.836&286.70230&-32.58164&2.43E-03&2.04E-03&53681.0831&-0.465266&-0.820736\\
1&31&F9D2370BM&159.836&286.70385&-32.58115&2.46E-03&2.28E-03&53681.0909&-0.465119&-0.820761\\
...&...&...&...&...&...&...&...&...&...&...
\enddata
\tablenotetext{a}{Set (orbit) number, star number (\#5 = \Her; reference star numbers same as Table~\ref{tbl-DR3p}),   OFAD-corrected X and Y positions in arcsec, position measurement errors in arcsec, time of observation = JD - 2400000.5, RA and DEC parallax factors. We provide a complete table in the electronic version of this paper. Set 18 from \HIP, \Gs DR1, and \Gs DR3; DR3 standard coordinates from Equations 1 and 2.}
\tablenotetext{b}{DR3 assumed oriented to RA, Dec. \HSTs orientation; spacecraft +V3 axis roll angle as defined in Chapter 2, \FGSs Instrument Handbook \citep{Nel15a}}
\tablenotetext{c}{\HSTs orbit and target identifier}
\tabletypesize{\normalsize}
\end{deluxetable}
\clearpage
\begin{deluxetable}{r l r r }
\tablewidth{0in}
\tablecaption{ Reference Star Parallax and Proper Motion Priors from \Gs DR3\label{tbl-pimu}}
\tablehead{
\colhead{ID}&
\colhead{$\varpi$ [mas]}&
\colhead{$\mu_{\rm RA}$ [\my]}&
\colhead{$\mu_{\rm Dec}$ [\my]}
}
\startdata
\Her&55.87$\pm$0.03&131.75$\pm$0.03&-297.03$\pm$0.04\\
30&0.22 0.01&-4.42 0.01&-10.37 0.02\\
31&7.63 0.01&-18.92 0.01&25.22 0.02\\
32&1.96 0.01&-0.26 0.01&-4.54 0.02\\
34&0.94 0.01&1.50 0.02&3.66 0.02\\
35&1.94 0.02&16.57 0.02&-26.78 0.02\\
61&2.82 0.18&0.58 0.19&0.72 0.21\\
62&3.07 0.01&-12.57 0.01&-0.92 0.01\\
63&0.67 0.01&-5.85 0.01&-2.03 0.02\\
65&6.21 0.01&-8.32 0.01&-29.79 0.01
\enddata
\end{deluxetable}
\clearpage

\begin{deluxetable}{l l}
\tablecaption{Reference Frame Statistics, \Hers Parallax, and Proper Motion\label{tbl-SUM}}
\tablewidth{6in}
\tablehead{\colhead{Parameter} &  \colhead{Value} }
\startdata
Study duration  &24.75 y  \\
number of observation sets    &   18  \\
reference star $\langle V\rangle$ &  12.79     \\ 
reference star $\langle (B-V) \rangle$ &0.81   \\
{\it HST}: model without \Her\,b, c orbits or \Hers DR3 priors \\ 
~~~~~~~$\chi^2/DOF$&0.617\\
~~~~~~~Absolute $\varpi$& 54.15 $\pm$ 0.22    mas \\
~~~~~~~Relative  $\mu_\alpha$& 132.84 $\pm$ 0.13 mas yr$^{-1}$\\
~~~~~~~Relative  $\mu_\delta$&  -297.60 $\pm$ 0.06  mas yr$^{-1}$\\
~~~~~~~$\vec{\mu} = 325.90$ mas  yr$^{-1}$\\
~~~~~~~P.A. = $155\fdg$95\\
{\it HST}: model with \Her\,b, c orbits, no \Hers DR3 priors\\ 
~~~~~~~$\chi^2/DOF$&0.463\\
~~~~~~~Absolute $\varpi$& 54.77 $\pm$ 0.22    mas \\
~~~~~~~Relative  $\mu_\alpha$& 132.44$\pm$ 0.2mas yr$^{-1}$\\
~~~~~~~Relative  $\mu_\delta$&  -298.00 $\pm$ 0.1  mas yr$^{-1}$\\
~~~~~~~$\vec{\mu} = 326.11$ mas  yr$^{-1}$\\
~~~~~~~P.A. = $156\fdg04$\\
\hline
DR3 catalog values\\
~~~~~~~Absolute $\varpi$& 55.87 $\pm$ 0.02     mas \\
~~~~~~~Absolute  $\mu_\alpha$& 131.75 $\pm$ 0.03 mas yr$^{-1}$\\
~~~~~~~Absolute  $\mu_\delta$&  -297.03 $\pm$ 0.04  mas yr$^{-1}$\\
~~~~~~~$\vec{\mu} = 324.93$ mas  yr$^{-1}$\\
~~~~~~~P.A. = $156\fdg08$
\enddata
\end{deluxetable}
\clearpage
\begin{deluxetable}{l r r}
\tablewidth{0in}
\tablecaption{14 Her\,{\lowercase{b, c}} Orbital Parameters from Astrometry and RV \label{ASTRVorb}}
\tablehead{
\colhead{Parameter} & 
\colhead{b} &
\colhead{c}
}
\startdata
$P$ [d]& 1767.56$\pm$ 0.22 & 52160$\pm$1028\\
$P$ [yr]& 4.8393$\pm$ 0.0006 & 142.8$\pm$2.8\\
$T$ [days] &51368.0 $\pm$ 0.5  & 51779$\pm$33\\
$e$  & 0.372$\pm$ 0.001 & 0.65$\pm$0.06 \\
$\omega$ [\arcdeg] &22.28$\pm$ 0.15 &  0 $\pm$ 1 \\ 
$K$ [\ms)] & 90.38 $\pm$ 0.15 & 50.8$\pm$ 0.4  \\
$\alpha$ [mas]& 1.28 $\pm$ 0.1 & 10.3$\pm$ 0.7 \\
$\Omega$ [$\arcdeg$]& 276 $\pm$ 5 & 224$\pm$ 9 \\
$i$ [$\arcdeg$]& 35.7 $\pm$ 3.2 & 82$\pm$14 \\
\hline 
Derived Parameters & &\\
$\alpha$ [AU]& 0.0233 $\pm$ 0.0019 & 0.187$\pm$ 0.012 \\
$a $ [AU] & 2.82 & 27\\
$a $ ["] &0.155 & 1.48\\
\msini [\mjupe] & 4.95 & 7.12\\
\m [\mjupe] & 8.5$^{+1.0}_{-0.8}$ & 7.1$^{+1.0}_{-0.6}$  \\
\m [\msune] & 0.0081 & 0.0068\\
\hline 
$\Phi$\tablenotemark{a}[$\arcdeg$]&~~~~~~62$\pm$ 12&
\enddata
\tablenotetext{a}{Mutual inclination from Equation~\ref{MI}}
\end{deluxetable}
\clearpage
\begin{deluxetable}{l  r r r r}
\tablecaption{KS Test Results\label{tbl-KSres}}
\tablewidth{4in}
\tablehead{
\colhead{Test} &  
\colhead{D} &
\colhead {C} &
\colhead {$\alpha$} &
\colhead {PV} 
		}
\startdata
\FGSs vs Feng22&0.33&0.39&0.050&0.090\\
Feng22 vs 6th Cat.&0.20&0.11&0.050&0.000\\
\FGSs vs 6th Cat.&0.44&0.38&0.050&0.005
\enddata
\end{deluxetable}

\clearpage

\clearpage



\clearpage

\end{document}